\documentstyle[aps,preprint,epsf]{revtex}
\begin{document}
\draft

\title{Reflection coefficient and localization length \\
of waves in one-dimensional random media}
\author{Kihong Kim\cite{kk}}
\address{Department of Physics, Ajou University, Suwon 442-749, Korea}
\maketitle 
\begin{abstract}
We develop a novel and powerful method of exactly calculating 
various transport characteristics
of waves in one-dimensional 
random media with (or without) coherent absorption or
amplification. Using the method, we compute 
the probability densities of the reflectance and of the phase of the reflection 
coefficient, together with the localization length, of electromagnetic waves
in sufficiently long random dielectric media. We find substantial differences
between our exact results and the previous results obtained using the random
phase approximation (RPA). The probabilty density of the phase of the reflection
coefficient is highly nonuniform when either disorder or absorption (or amplification) is strong. 
The probability density of the reflectance when the absorption
or amplification parameter is large is also quite different from the RPA result.
We prove that the probability densities 
in the amplifying case are  
related to those in the 
absorbing case with the same magnitude of the imaginary part of the 
dielectric permeability by exact dual relationships.
From the analysis of the average reflectance 
that shows a nonmonotonic dependence on the 
absorption or amplification parameter, we obtain a useful criterion for the applicability 
of the RPA. In the parameter regime where the RPA is invalid, we find 
the exact localization length is substantially
larger than the RPA localization length.
\\ 

\pacs{PACS Numbers: 42.25.Bs, 72.15.Rn, 73.20.Dx, 78.20.Ci}
\end{abstract}


\section{Introduction}
\label{sec-int}

Propagation and localization of waves of various kinds in 
one-dimensional disordered media
have been studied intensively for several decades.
Examples of particular interest are {\it classical} electromagnetic wave
propagation in random dielectric media \cite{ping,frei} and
{\it quantum} electron transport in disordered solids \cite{abri,lee,lgp}.
Much recent attention has been paid to the problem of
wave propagation in coherently absorbing
or amplifying random media 
\cite{fpy,jay1,fp,pk,zyu,bpb,mpb,fpy2,sen2,jay2}. 
It has been demonstrated that the wave is more strongly localized in
both types of media than in unitary (or elastic) random media 
with no absorption and amplification.

A large number of previous studies have been 
based on the so-called random phase approximation (RPA),
where it is assumed that the phase of the reflected 
wave relative to that of the incident wave is uniformly
and randomly distributed over all angles.
One can easily prove that this approximation is 
equivalent to assuming large wave energy or
weak disorder, and, at the same time, 
weak absorption or amplification. 
When these conditions are not met, an exact calculation 
beyond the random phase approximation is required.

Recently, several authors have performed numerical calculation
of the probability densities of the transmittance, the reflectance, 
and the phase of the complex reflection coefficient
in absorbing, amplifying and unitary random media
using methods that goes beyond the random phase approximation 
\cite{jay1,sen2,jay2,sen1,prad1}.
Their results clearly indicate that the random phase approximation
can often lead to qualitatively wrong conclusions on the behavior
of waves in random media.
Above all, it has been demonstrated that the distribution of the phase
of the reflection coefficient is generally nonuniform in all of absorbing,
amplifying and unitary media.
The distribution of the reflectance is also quite different from
the RPA result. For example, the average
reflectance in the RPA is a monotonically decreasing function of 
the absorption parameter, whereas the exact
average reflectance is a nonmonotonic one \cite{jay1}.
 
In the present work, we develop a novel and powerful method of 
$exactly$ calculating various disorder-averaged quantities 
including the reflectance, the transmittance,
and their probability densities
in absorbing, amplifying and unitary media.
We also calculate the probability density of the phase of the reflection 
coefficient and the localization length exactly. In addition, we derive a 
couple of exact dual relationships that relate the probability densities in
the absorbing case to those in the amplifying case with the same magnitude
of the imaginary part of the dielectric permeability. 
From the quantitative analysis
of the average reflectance in the absorbing case, we obtain 
a useful criterion for the applicability of the RPA.

The outline of the paper is as follows.
In the next section, we introduce the model.
In Sec.~\ref{sec-met}, we describe our method 
of calculating the probability densities 
and the localization length in detail. We also prove the exact dual
relationships that relate the probability densities in the amplifying
case to those in the absorbing case.
Results of our calculation are presented in Sec.~\ref{sec-res}.
Finally, we summarize the paper in Sec.~\ref{sec-conc}.

\section{Model}
\label{sec-model}

We are interested in the propagation of 
a monochromatic electromagnetic wave of frequency $\omega$ 
in disordered dielectric media. For the sake of 
simplicity, we consider
the one-dimensional case, where the dielectric permeability $\epsilon$
varies only in one direction in space and the wave
propagates in the same direction. We take this
direction as the $z$-axis and assume the medium lies in  
$0 \le z \le L$. Then the complex amplitude of the electric field, $E$,
satisfies the Helmholtz equation with the wavenumber 
$k=\omega/c$, where $c$ is the speed of light in a vacuum,
\begin{equation}
\left[ {{d^2}\over{dz^2}}+k^2\epsilon(z)\right] E(z)=0.
\end{equation}
The dielectric permeability $\epsilon(z)$ is equal to 1
for $z<0$ and $z>L$ and $\epsilon(z)=1+a+\delta\epsilon(z)
+i\gamma$ for $0 \le z \le L$,
where $a$ and $\gamma$ are assumed to
be real constants and $\delta\epsilon(z)$ is 
a real random function of $z$.
We suppose $\delta\epsilon(z)$ to be a Gaussian random function with 
zero mean and a white-noise spectrum:
\begin{equation}
\langle \delta\epsilon(z)\delta\epsilon(z^\prime) \rangle = 
g\delta(z-z^\prime),~~~\langle\delta\epsilon(z)\rangle=0,
\end{equation}
where $\langle\cdots\rangle$ denotes an average over disorder and
the constant $g$ is a measure of the strength of randomness.
The imaginary part of the dielectric permeability, $\gamma$,
makes the medium absorb ($\gamma>0$) or
amplify ($\gamma<0$) the wave without destroying its 
phase coherence. For simplicity, $\gamma$ is assumed to be uniform.
The constant $1+a$ is the disorder average of the real part of the 
dielectric permeability and, in general, is not equal to 1.
Our method is applicable to the cases where $a$ is an arbitrary
nonrandom function, e.g. a periodic function, of $z$, 
but in this
paper, we restrict our attention to the case where $a$ 
is a constant. 

The above model is also relevant to the electron transport
problem in disordered quasi-one-dimensional 
solids. In that case, the electron
wave function $\psi(z)$ plays the same role as $E(z)$
and $\epsilon(z)$ is replaced by $1-U(z)/E_0$, where
$U(z)$ is the random potential experienced by electrons
and $E_0$ is the kinetic energy of incident electrons.

\section{Method}
\label{sec-met}

We consider a plane wave of unit magnitude $E(z)=e^{ik(L-z)}$ 
incident on the medium from the right. The quantities of main interest
are the complex reflection and transmission coefficients 
$r=r(L)$ and $t=t(L)$ defined
by the wave function outside the medium:
\begin{equation}
E(z)=\left\{ \begin{array}{ll}
e^{ik(L-z)}+r(L)e^{ik(z-L)},  &  ~z>L, \\
t(L)e^{-ikz},  &  ~z<0.  \end{array} \right.
\end{equation}
Using the so-called {\it invariant imbedding} method \cite{dou,klya}, we derive exact 
differential equations for $r$ and $t$ with respect to $L$:
\begin{eqnarray}
{{dr(L)}\over{dL}}&=&2ikr(L)+{{ik}\over 2}\left[a+i\gamma+\delta\epsilon(L)
\right]\left[1+r(L)\right]^2,~~~r(L=0)=0 \label{eq:r},\\
{{dt(L)}\over{dL}}&=&ikt(L)+{{ik}\over 2}\left[a+i\gamma+\delta\epsilon(L)
\right]\left[1+r(L)\right]t(L),~~~t(L=0)=1 \label{eq:t}.
\end{eqnarray}                 
These stochastic differential equations can be used in calculating 
the disorder averages of various physical quantities consisted of
$r$ ($\equiv\sqrt{R}e^{i\theta}$) and $t$ ($\equiv\sqrt{T}e^{i\phi}$), 
where the reflectance $R=\vert r\vert^2$ and the transmittance 
$T=\vert t\vert^2$ as well as the phases $\theta$ and $\phi$ are
functions of $L$.
In the present work, we are mainly interested in computing the 
{\it exact} probability densities 
$P_R(R)$ and $P_{\theta}(\theta)$
in semi-infinite ($L\rightarrow\infty$) absorbing and amplifying 
random dielectric media.
We will also compute the {\it exact} localization length $\xi$ 
of the wave in absorbing random media defined by 
\begin{equation}
\lim_{L\rightarrow\infty}\langle \log T\rangle=-L/\xi.
\label{eq:ll}
\end{equation}
The unitary case with no absorption or amplification ($\gamma=0$)
is also of great interest. In that case, the probability density
$P_R(R)$ in the $L\rightarrow\infty$ limit is clearly
equal to $\delta(R-1)$ regardless of the strength of 
disorder $g$. But the probability density $P_\theta(\theta)$ and the 
localization length $\xi$ depend on $g$ in a nontrivial manner
and will be studied in the present work.

\subsection{Probability Distribution of the Reflectance}
\label{sec-ref}

We obtain the probability density 
$P_R(R)$ from the moments
$\langle R^n\rangle$ 
for all integers $n$. An infinite number of
coupled non-stochastic ordinary differential equations 
satisfied by these moments are obtained using Eq.~(\ref{eq:r}) and
Novikov's formula \cite{nov}. It turns out that in order to compute 
$\langle R^n\rangle=\langle r^n {r^*}^n\rangle$, 
one needs to compute the moments 
$Z_{n\tilde n}\equiv \langle r^n{r^*}^{\tilde n}\rangle$
for all integers $n$ and $\tilde n$.
In other words, the moments $Z_{n\tilde n}$ with $n=\tilde n$
are coupled to $Z_{n\tilde n}$ with $n\ne\tilde n$.
Using the notations $l=L/\xi_0$, $C=k\xi_0$, $\alpha=k\xi_0 a$
and $\beta=k\xi_0\gamma$, where $\xi_0=4/gk^2$ is the 
localization length with no absorption or amplification
in the random phase approximation,
we obtain the {\it nonrandom} equation satisfied by $Z_{n\tilde n}$:
\begin{eqnarray}
{{dZ_{n\tilde n}}\over{dl}}&=&\left[i(2C+\alpha)(n-\tilde n)
- \beta(n+\tilde n)
+4n\tilde n-3n^2-3{\tilde n}^2\right] Z_{n\tilde n}   \nonumber\\
&+& n\left[ {1 \over 2}(i\alpha-\beta)
-2n+2\tilde n-1\right]Z_{n+1,\tilde n}   
+ n\left[ {1 \over 2}(i\alpha-\beta)
-2n+2\tilde n+1\right]Z_{n-1,\tilde n}  \nonumber\\
&-& \tilde n\left[ {1 \over 2}(i\alpha+\beta)
-2n+2\tilde n+1\right]Z_{n,\tilde n+1} 
- \tilde n\left[ {1 \over 2}(i\alpha+\beta)
-2n+2\tilde n-1\right]Z_{n,\tilde n-1}  \nonumber\\
&+& n\tilde n Z_{n+1,\tilde n+1}+n\tilde n Z_{n-1,\tilde n-1}
+n\tilde n Z_{n+1,\tilde n-1}+n\tilde n Z_{n-1,\tilde n+1}  \nonumber\\
&-& {1 \over 2} n(n+1) Z_{n+2,\tilde n}
-{1\over 2}n(n-1)Z_{n-2,\tilde n} 
- {1 \over 2} \tilde n(\tilde n+1) Z_{n,\tilde n+2}
-{1\over 2}\tilde n(\tilde n-1)Z_{n,\tilde n-2},
\label{eq:z} 
\end{eqnarray}
with the conditions $Z_{00}=1$ and $Z_{n\tilde n}(l=0)=0$
for $n,~\tilde n>0$.
The random phase approximation
applies to the case
where $C\gg 1$ and $C\gg\beta$, that is $\gamma\ll 1$. Then
it is clear that $Z_{n\tilde n}$ with $n\ne\tilde n$
can be neglected. This leads to the much simpler equation 
for $Z_n\equiv Z_{nn}=\langle R^n\rangle$ obtained
previously in \cite{fpy}:
\begin{equation}
{{dZ_n}\over{dl}}=-2n(n+\beta)Z_n+n^2Z_{n+1}+n^2Z_{n-1},
\end{equation}
with the conditions $Z_0=1$ and $Z_{n>0}(l=0)=0$.
In this paper, we go beyond the random phase approximation 
and solve the exact equation (\ref{eq:z}) directly.
  
We consider the absorbing ($\beta>0$) 
case first.
When $\beta$ is positive, the moments $Z_{n\tilde n}$ with $n,~\tilde
n\ge 0$ are coupled to one another and well-behaved for all $l$.
Furthermore, the magnitude of the moment $Z_{n\tilde n}$
decays (more rapidly for larger $\beta$ values) as
either $n$ or $\tilde n$ increases.
Based on this crucial observation, 
we solve the infinite number of coupled
equations, Eq.~(\ref{eq:z}), by a simple truncation method,
which was developed and applied successfully
to the problem of computing the electronic properties
of quasi-one-dimensional Peierls systems by the author and collaborators in previous publications \cite{kim1,kim2}. We assume
$Z_{n\tilde n}=0$ for either $n$ or $\tilde n$ greater than some large
positive integer $N$ and solve the {\it finite} number ($=N(N+2)$) of 
coupled equations
numerically for given values of $C$, $\alpha$ and $\beta$. 
We increase the cutoff $N$, repeat a similar calculation,
and then compare the newly-obtained $Z_{n\tilde n}$ 
with the value of the previous step. If there is no
change in the values of $Z_{n\tilde n}$ within 
an allowed numerical error, we conclude that 
we have obtained the exact solution of $Z_{n\tilde n}$.
In the present work, we will limit our attention further
and consider only the $l\rightarrow\infty$ limit, where
we expect $dZ_{n\tilde n}/dl=0$. Then the left-hand side 
of Eq.~(\ref{eq:z}) vanishes and we have a set  
of coupled {\it algebraic} equations. 
We solve these equations by the truncation method described above
to find $\langle R^n\rangle$ for every integer $n>0$.

It is possible to get the probability density $P_R(R)$
from the moments by several different methods. 
We find it is efficient to use the expansion of $P_R(R)$
in terms of the shifted Legendre polynomials:
\begin{equation}
P_R(R)=\sum_{m=0}^\infty (2m+1)\langle P_m^*(R)\rangle
P_m^*(R),
\end{equation}
where $P_m^*(R)$ is the shifted Legendre polynomial
of order $m$ defined over the interval $0\le R\le 1$.
The average value $\langle P_m^*(R)\rangle$ 
is computed using the moments $\langle R^n\rangle$ for $0 \le n \le m$ 
and turns out to be a rapidly decreasing function of $m$.

The amplifying ($\beta<0$) case is a little trickier.
Our method fails since $\vert Z_{n\tilde n}\vert$
does not decay as $n$ or $\tilde n$ increases to large 
positive values. For sufficiently large values of $l$, however,
Eq.~(\ref{eq:z}) is well-defined if $n$ and $\tilde n$ are
negative. From Eq.~(\ref{eq:z}), we can easily prove that 
$Z_{n\tilde n}(\beta=\beta_0)=Z_{-\tilde n,-n}(\beta=-\beta_0)$.
Thus we obtain
\begin{equation}
\langle R^n\rangle_{\beta=\beta_0}=
\langle R^{-n}\rangle_{\beta=-\beta_0},
\end{equation}                         
which is equivalent to  
\begin{equation}
P_R(R,\beta=-\beta_0)={1 \over {R^2}} P_R\left({1 \over R},
\beta=\beta_0\right).       
\label{eq:dual}
\end{equation}
We observe that 
Eq.~(\ref{eq:dual}) suggests $P_R(R)$ to be nonzero  
only for $1 \le R \le \infty$ in the amplifying case.
This result does not make much sense for $l \ll 1$
because in that case, we expect  
$P_R(R)$ for $\beta<0$ as well as 
$P_R(R)$ for $\beta>0$ to have a sharp peak around $R=0$.
Nevertheless, we conjecture that Eq.~(\ref{eq:dual}) is
true for all $l \gg 1$, that is for all $L\gg\xi_0$. 
Once we get $P_R(R)$ for $\beta<0$,
we can compute the moments
$\langle R^n\rangle$ using the definition
\begin{equation}
\langle R^n\rangle_{\beta<0} 
=\int_1^\infty dR~R^n P_R(R,\beta<0).
\end{equation}

The unitary case requires a separate consideration. We
expect all $Z_{n\tilde n}$'s with $n=\tilde n$ are 
equal to 1 and $P_R(R,\beta=0)=\delta(R-1)$ in the $l\rightarrow\infty$ limit.
Therefore we have to set $Z_{nn}$ for $n=N+1$ 
equal to 1 instead of 0 in solving Eq.~(\ref{eq:z}).
This changes the right-hand side of the last one of our 
$N(N+2)$ algebraic equations, by solving which 
we obtain all moments of the form 
$\langle e^{i(n-\tilde n)\theta}\rangle$. The result  
will be used in Secs.~\ref{sec-ph} and \ref{sec-loc} in 
calculating the probability density of the 
phase of the reflection coefficient and the localization 
length exactly.

\subsection{Probability Distribution of the Phase of 
the Reflection Coefficient}
\label{sec-ph}

The probability density $P_\theta(\theta)$ is most easily
obtained using the Fourier series expansion
\begin{equation}
P_\theta(\theta)={1 \over {2\pi}}+{1 \over \pi}\sum_{m=1}^\infty
\langle\cos(m\theta)\rangle\cos(m\theta)
+ {1 \over \pi}\sum_{m=1}^\infty
\langle\sin(m\theta)\rangle\sin(m\theta).
\end{equation}
Therefore we need the averages $\langle\cos(m\theta)\rangle$
and $\langle\sin(m\theta)\rangle$ for every positive integer $m$.
Unfortunately, we are not aware of how to compute these averages
directly when $m$ is an odd integer 
unless $\beta=0$. 
For even integers $m=2n$, we can calculate
$\langle e^{2in\theta}\rangle=\langle \cos(2n\theta)\rangle
+i\langle\sin(2n\theta)\rangle$ using the relation
$\langle e^{2in\theta}\rangle=Z_{n,-n}=\langle (r/r^*)^n\rangle$.
Since $Z_{n,-n}$ is coupled to every $Z_{n\tilde n}$ with $n\ge 0$
and $\tilde n\le 0$, we need to calculate $Y_{n\tilde n}
\equiv Z_{n,-\tilde n}$ with $n,~\tilde n \ge 0$.
The equation satisfied by $Y_{n\tilde n}$ is obtained from
Eq.~(\ref{eq:z}) by a simple substitutuion:
\begin{eqnarray}
{{dY_{n\tilde n}}\over{dl}}&=&\left[i(2C+\alpha)(n+\tilde n)
- \beta(n-\tilde n)
-4n\tilde n-3n^2-3{\tilde n}^2\right] Y_{n\tilde n}   \nonumber\\
&+& n\left[ {1 \over 2}(i\alpha-\beta)
-2n-2\tilde n-1\right]Y_{n+1,\tilde n}   
+ n\left[ {1 \over 2}(i\alpha-\beta)
-2n-2\tilde n+1\right]Y_{n-1,\tilde n}  \nonumber\\
&+& \tilde n\left[ {1 \over 2}(i\alpha+\beta)
-2n-2\tilde n-1\right]Y_{n,\tilde n+1} 
+ \tilde n\left[ {1 \over 2}(i\alpha+\beta)
-2n-2\tilde n+1\right]Y_{n,\tilde n-1}  \nonumber\\
&-& n\tilde n Y_{n+1,\tilde n+1}-n\tilde n Y_{n-1,\tilde n-1}
-n\tilde n Y_{n+1,\tilde n-1}-n\tilde n Y_{n-1,\tilde n+1}  \nonumber\\
&-& {1 \over 2} n(n+1) Y_{n+2,\tilde n}
-{1\over 2}n(n-1)Y_{n-2,\tilde n} 
- {1 \over 2} \tilde n(\tilde n+1) Y_{n,\tilde n+2}
-{1\over 2}\tilde n(\tilde n-1)Y_{n,\tilde n-2},
\label{eq:y} 
\end{eqnarray}
with the condition $Y_{00}=1$. 
We find that when $C\gg 1,~\beta$ or more precisely,
when $\vert 2C+\alpha\vert\gg 1,~\beta$, 
all $Y_{n\tilde n}$'s except $Y_{00}$ are negligible,
which obviously means the phase $\theta$ is 
distributed randomly and uniformly
 over the interval $0 \le \theta\le 2\pi$.
We solve Eq.~(\ref{eq:y}) in the $l\rightarrow \infty$ limit
for $\beta>0$ using the truncation 
method described in the previous section. 
In order to obtain $P_\theta(\theta)$,
we also need to find $\langle e^{im\theta}\rangle$ for odd $m$.
For this purpose, we make a conjecture 
that $\langle e^{im(\theta-\pi)}\rangle$
is a {\it smooth} function of $m$, the validity of which
can be tested directly in the unitary case 
where we can compute $\langle e^{im\theta}\rangle$ 
for every integer $m$. Then the averages 
$\langle e^{im\theta}\rangle$ for odd $m$ can be obtained 
by a numerical interpolation between the averages for even $m$. 
Once we get $P_\theta(\theta)$ for $\beta>0$,
it is trivial to find $P_\theta(\theta)$ for $\beta<0$. 
We easily see from
Eq.~(\ref{eq:y}) that $Y_{n\tilde n}(\beta=\beta_0)
=Y_{\tilde n n}(\beta=-\beta_0)$. This implies
\begin{equation} 
P_\theta(\theta,\beta=\beta_0)=P_\theta(\theta,\beta=-\beta_0).
\end{equation}

\subsection{Localization Length}
\label{sec-loc}

In order to compute the localization length
as defined by Eq.~(\ref{eq:ll}), we need to compute
the average $W\equiv \langle \log T\rangle$ in the $l\rightarrow
\infty$ limit. The nonrandom differential 
equation satisfied by $W$ in the absorbing ($\beta>0$) 
and unitary ($\beta=0$) cases is obtained
using Eqs.~(\ref{eq:r},\ref{eq:t}) and Novikov's formula in a 
straightforward manner:
\begin{equation}
{{dW}\over{dl}}=-(1+\beta)-{\rm Re}\left[(2+\beta-i\alpha)Z_{10}
+Z_{20}\right].
\end{equation}
In case of the random phase approximation, 
$Z_{10}$ and $Z_{20}$ vanish 
and the RPA localization length becomes \cite{fpy}
\begin{equation}
\xi_{\rm RPA}={{\xi_0}\over {1+\beta}}.
\end{equation}                         
$Z_{10}$ and $Z_{20}$, however, do not vanish in general and
the exact localization length is obtained from
\begin{equation}
{\xi_0 \over \xi}=1+\beta
+{\rm Re}\left[(2+\beta-i\alpha)Z_{10}(l\rightarrow\infty)
+Z_{20}(l\rightarrow\infty)\right],
\label{eq:ell}
\end{equation}
where the asymptotic values of $Z_{10}$ and $Z_{20}$ 
found in Sec.~\ref{sec-ref} are used.

\section{Results}
\label{sec-res}

All of our results were obtained for the cutoff $N=60$. In other words,
we have solved $60\times 62=3720$ coupled algebraic equations
numerically. We have confirmed that the
results presented in this section are exact for all practical
purposes.

\subsection{Probability Distribution of the Reflectance}

We consider the absorbing case first.
Fig. 1(a) shows the probability density of the reflectance
$P_R(R)$ in the large distance limit for $C=5$, $\alpha=0$
and $\beta=1,3,6,10,20$. Fig. 1(b) is the
probability density of the reflectance for the same $\beta$ values 
in the random phase approximation (that is, $C=\infty$), the analytical
form of which was obtained previously in \cite{pk}:
\begin{equation}
P_R(R,C=\infty)={{2\beta\exp\left(-{{2\beta R}\over{1-R}}\right)}
\over{\left(1-R\right)^2}}  ~~~~{\rm for}~~ 0\le R\le 1.
\end{equation} 
For small $\beta$ values ($\beta=1,3$), 
we find that the exact $P_R(R)$ agrees pretty well with the RPA
probability density. As $\beta$ increases, however, there
appear remarkable differences between the exact and RPA probability
densities. Most
notably, the exact $P_R(R)$ develops a sharp peak at nonzero 
$R=R_{\rm max}$, while the RPA probability density has a peak
at $R=0$ for all $\beta\ge 1$. 
This $R_{\rm max}$ increases and the half-width
of the peak decreases as $\beta$ increases further.
We have observed similar behavior for other values of $C$.
In the parameter range we have explored in detail 
($1\le C,\beta\le 10$, $\alpha=0$), 
we do not find a double-peaked structure
in the exact probability density reported in \cite{jay1}.  

In Fig. 2, we show the average reflectance $\langle R\rangle$ 
as a function of $\beta$ for $C=1,5,10,\infty$ and
$\alpha=0$. In the RPA case, $\langle R\rangle$ is a 
monotonically decreasing function of $\beta$. Since $1-\langle R\rangle$
is the amount of average absorption, 
this means that absorption increases monotonically 
as the dimensionless absorption parameter $\beta$ gets bigger. Though this
may sound reasonable, it is actually a false conclusion. As pointed out in
\cite{jay1}, the medium with a sufficiently large 
$\beta$ behaves as a 
reflector rather than as an absorber. In agreement with \cite{jay1}, 
we have found that $\langle R\rangle$ reaches a minimum at 
$\beta=\beta_{\rm min}$ and is an increasing function of $\beta$ for 
$\beta>\beta_{\rm min}$. 
We have checked numerically that 
 in the $\beta\rightarrow\infty$ limit, $\langle R\rangle
\rightarrow 1$ and $P_R(R)\rightarrow
\delta(R-1)$.
As is obvious from Fig. 2, $\beta_{\rm min}$ increases as $C$
increases and can be used as a useful criterion for the region of 
validity of the random phase approximation \cite{jay1,fp}. 
That is, the RPA is approximately valid when $\vert\beta\vert<\beta_{\rm min}$
(and $C\gg 1$ (see Sec.~\ref{sec-phd})).
In Fig. 3, we plot $\beta_{\rm min}$
versus $C$ for $\alpha=0$. It is a monotonically increasing function
and is fitted fairly well by a power law function $\beta_{\rm min}
=aC^b$ with $a\approx 2.04 \pm 0.03$ and $b\approx 0.58\pm 0.01$.

Next we show the probability density of the relectance in the amplifying
case, which we obtain quite easily using the dual relationship Eq.~(\ref{eq:dual}).
In Fig. 4(a), we show $P_R(R)$ for $C=5$, $\alpha=0$, 
and $\beta=-1,-3,-6,-10,-20$. Fig. 4(b) is the RPA probability density
for the same $\alpha$ and $\beta$ values. In the RPA case, $P_R(R)$ is
proportional to $1/R^2$ in the $R\rightarrow \infty$ limit for all $\beta<0$,
since $P_R(R=0)$ is a finite constant for all $\beta>0$. 
This implies the average
reflectance in the amplifying case is always divergent. 
The RPA result is wrong, however, because when
$\beta\rightarrow -\infty$, the medium has to behave as a pure reflector with
$\langle R\rangle =1$. In Fig. 1(a), we observe that $P_R(R)$ is finite
at $R=0$ for $\beta=1,3,6$, but goes to zero as $R\rightarrow 0$ 
for sufficiently large $\beta$ values. This observation and the dual relationship 
Eq.~(\ref{eq:dual}) ensures that $\langle R\rangle$ is finite for sufficiently
large negative $\beta$'s.

Finally, in Figs. 5(a) and 5(b), we show the probability density of the 
reflectance for $\beta=5,-5$, $\alpha=0$ and $C=1,2,4,7,\infty$. In this way,
we can clearly see how the exact $P_R(R)$ departs from the RPA result as $C$
decreases.

\subsection{Probability Distribution of the Phase of 
the Reflection Coefficient}
\label{sec-phd}

As explained in Sec.~\ref{sec-ph}, we have difficulty in obtaining the exact
probability distribution of the phase of the reflection coefficient 
$P_\theta(\theta)$ for 
arbitrary $\beta$ values in the absorbing and amplifying cases. 
This difficulty does not arise in the unitary case ($\beta=0$). Therefore we 
show the probability density in this case first 
in Figs. 6(a-c) for a wide range of $C$ values
and $\alpha=0$. For sufficiently small $C$'s, $P_\theta(\theta)$ has a sharp 
symmetric peak located at $\theta=\pi$. 
We expect $P_\theta(\theta)$ approaches
$\delta(\theta-\pi)$ as $C\rightarrow 0$.
As $C$ increases from zero, 
this large peak moves to $\theta$ slightly bigger than
$\pi$ and  
a small secondary peak appears at $\theta$ smaller than $\pi$.
As $C$ is increased further, 
the large peak keeps moving away from $\theta=\pi$ 
and another small secondary peak
is developed at $\theta<\pi$. The overall shape 
of the probability density becomes broader.
When $C\approx 0.05$, the small peaks merge and are turned into a flat region.  
For $C \gg 1$, $P_\theta(\theta)$ is almost constant with a small 
and broad peak at $\theta=5\pi/3$ and a valley at $\theta=\pi/3$.

In the absorbing and amplifying cases, 
we can calculate $P_\theta(\theta)$ reliably
only when $\langle e^{im(\theta-\pi)}\rangle$ is a smooth
function of the integer $m$.
It turns out that our interpolation method described 
in Sec.~\ref{sec-ph} works when 
$\vert\beta\vert$ is sufficiently large compared to $C$ or $C$ is 
sufficiently small, in other words, 
when the random phase approximation does not work. 
In Fig. 7, we illustrate the behavior of $P_\theta(\theta)$ 
for a rather small value of $C$ ($=0.1$) and $\alpha=0$, 
$\vert\beta\vert=0.1,0.6,1,1.5,2,3$. 
When $\vert\beta\vert\gg C$, 
$P_\theta(\theta)$ has a sharp peak at $\theta=\pi$.
As $\vert\beta\vert$ decreases, this peak becomes lower and broader.
When $\beta\stackrel{<}{\sim}2$, 
the peak shifts to $\theta<\pi$ and a new peak appears
at $\theta>\pi$. This new peak grows and the old peak decays as $\vert\beta\vert$
increases further. At $\beta=0.1$, $P_\theta(\theta)$ is 
almost identical to the probability density in the unitary case (Fig. 6(b)).

\subsection{Localization Length}

In Fig. 8(a), we plot the inverse (dimensionless) 
localization length $\xi_0/\xi$ in the unitary case 
versus $\log_{10}C$. When $C\gg 1$, 
$\xi$ is close to the RPA localization length in the unitary case 
 $\xi_0$. In the opposite limit $C\rightarrow 0$,
$\xi_0/\xi$ goes to zero, or equivalently $\xi/\xi_0$ {\it diverges}.
 Fig. 8(b) shows the same data on a log-log plot.
We note that the $C\ll 1$ region is approximately linear and is fitted by a 
power law function $\xi_0/\xi =a^\prime C^{b^\prime}$ with $a^\prime
\approx 0.81$ and $b^\prime\approx 0.65$. 

Finally, in Fig. 9(a), we show the localization length in the absrbing case
as a function of $\beta$ for $C=5$ and $C=\infty$. We note that 
the exact localization length is always larger than the RPA 
localization length ($=\xi_0/(1+\beta)$) 
for the same $\beta$. Fig. 9(b) shows the localization
length as a function of $C$ for $\beta=5$.

\section{Conclusion}
\label{sec-conc}

In this paper, we have presented a novel numerical method of 
calculating various transport characteristics of waves in 
one-dimensional random media with (or without) 
absorption or amplification and used it to obtain the 
probability densities of the reflectance 
and of the phase of the reflection
coefficient in the large distance limit, 
together with the localization length of waves. Our method is completely
beyond the random phase approximation that has been used frequently 
in previous works and gives essentially exact results in the sense that the 
numerical error is unnoticeably small in all of the figures presented in 
this paper. 
The probability distribution of the phase of the reflection coefficient 
turns out to be highly nonuniform 
when either the disorder parameter or the absorption (or amplification)
parameter (that is, 
the magnitude of the imaginary part of the dielectric permeability, $\vert\gamma\vert$)
is large. When the absorption or amplification parameter is large, the 
probability distribution of the reflectance shows behavior that is
totally different from the RPA behavior.  
We have also proved a couple of 
exact dual relationships between 
the probability densities in the absorbing 
case and those in the amplifying case with the same $\vert\gamma\vert$.
The probability density of the phase of the reflection coefficient
in the amplifying case is the same as that in the absorbing case with the 
same $\vert\gamma\vert$. The probability density of the reflectance is
obtained from that in the absorbing case with the same $\vert\gamma\vert$
by a simple transformation (Eq.~(\ref{eq:dual})). 
From the quantitative analysis of the average reflectance that shows a 
nonmonotonic dependence on the absorption or amplification parameter, we 
find a criterion for the applicability of the RPA. In the 
parameter regime where the RPA
is invalid, we find the exact localization length is much larger than the RPA
localization length.

In the present work, we have limited our attention to the large distance limit. However, we
could have integrated the differential equations Eqs.~(\ref{eq:z}) and (\ref{eq:y})
directly to obtain the probability densities of the finite-size system.
Our method can also be generalized in a straightforward manner 
to the calculation of the probability 
densities of the transmittance and the phase of the tranmission coefficient. 
Work in this direction is in progress and will be presented elsewhere.
  
\ \

\centerline{\bf ACKNOWLEDGMENTS}

\ \
The author is grateful to P. Pradhan for sending him a copy of 
the reference \cite{prad1}.
This work has been supported by the Korea Science and Engineering Foundation
through grant number 95-0701-02-01-3.

\begin{figure}
\protect\centerline{\epsfxsize=5in \epsfbox{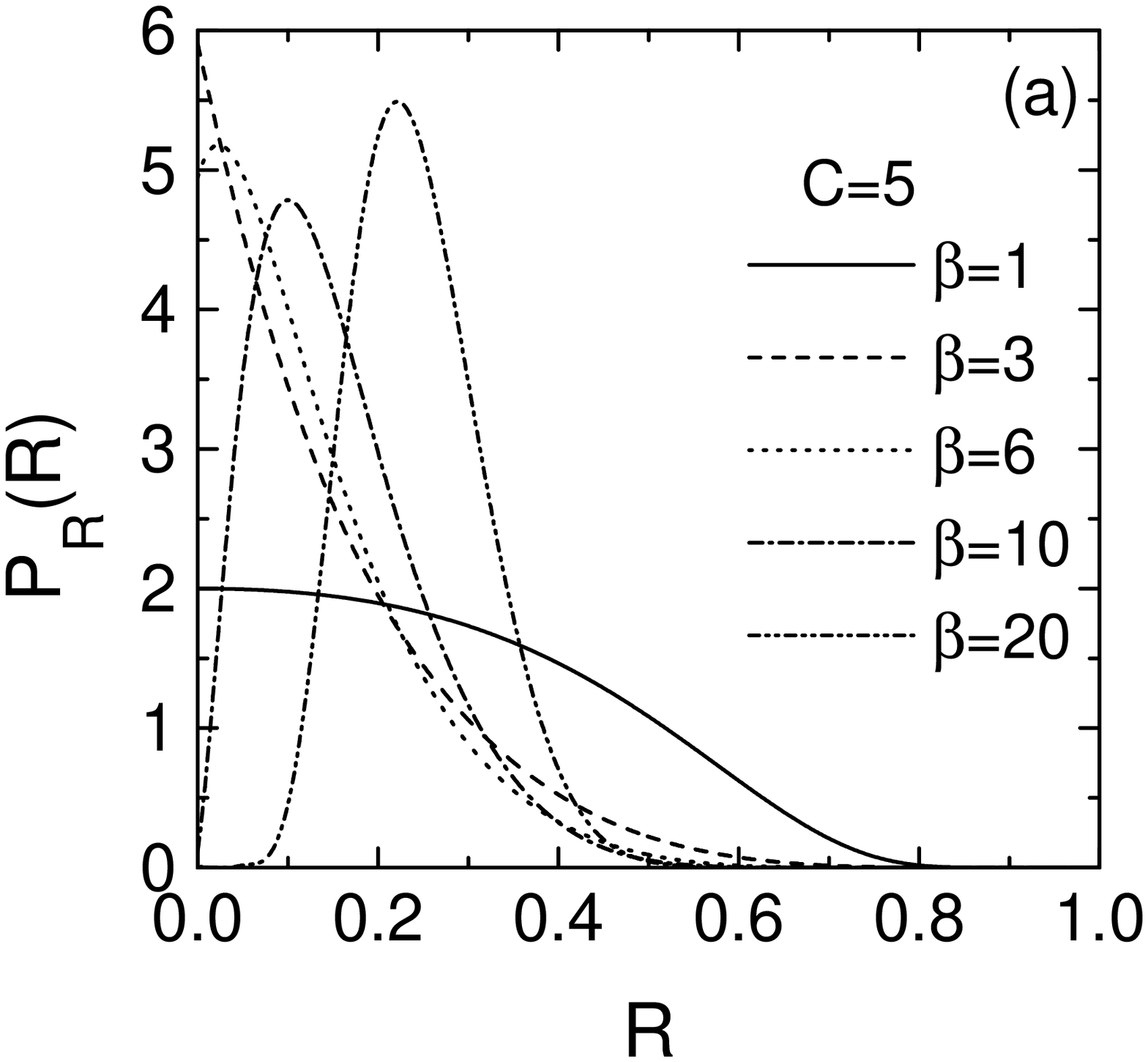}}
\protect\centerline{\epsfxsize=5in \epsfbox{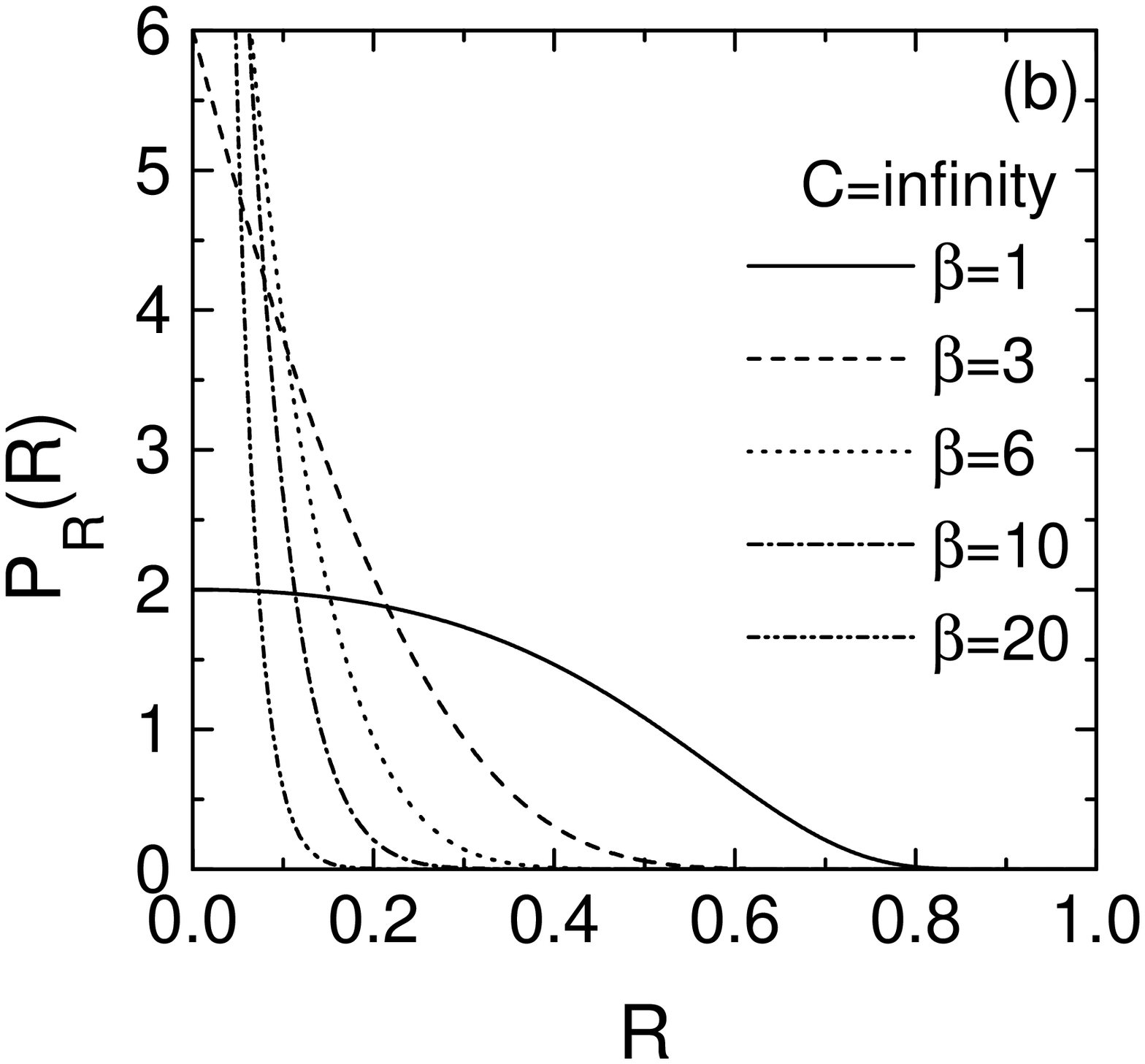}}
\caption{Probability density of the reflectance in the absorbing case 
for $\alpha=0$, $\beta=1,3,6,10,20$ and (a) $C=5$ (exact result) or  
(b) $C=\infty$ (RPA).}
\end{figure}

\begin{figure}
\protect\centerline{\epsfxsize=5in \epsfbox{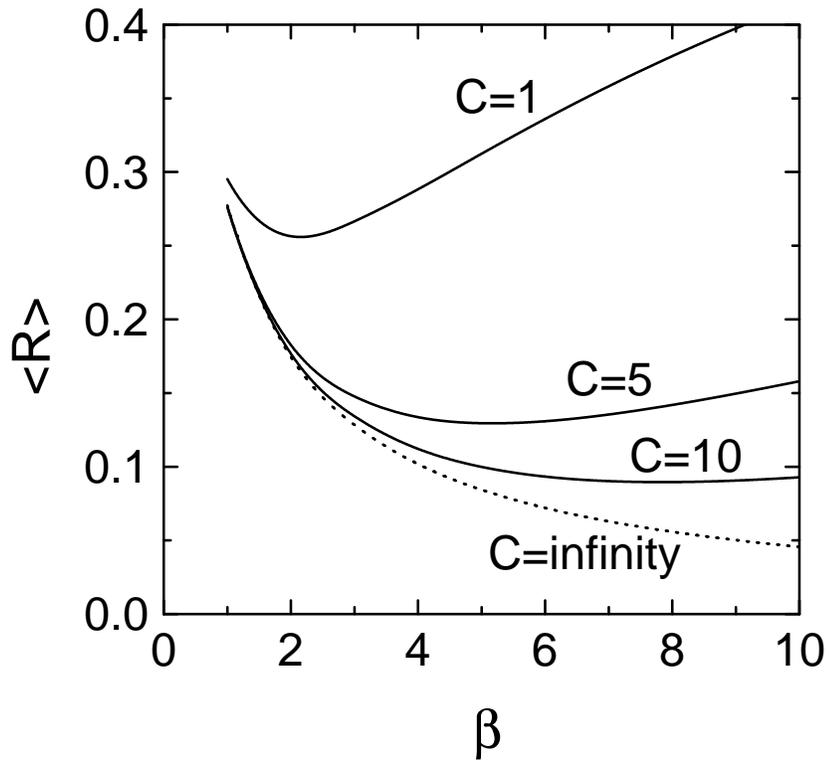}}
\caption{Average reflectance versus $\beta$ in the absorbing case for
$\alpha=0$ and $C=1,5,10,\infty$.}
\end{figure}

\begin{figure}
\protect\centerline{\epsfxsize=5in \epsfbox{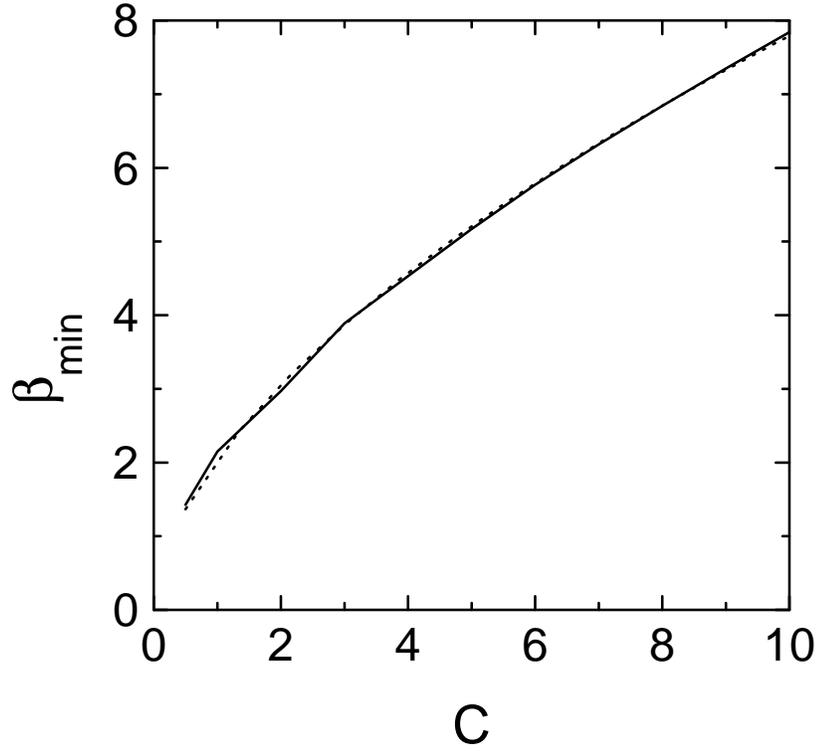}}
\caption{$\beta_{\rm min}$ versus $C$ for $\alpha=0$. $\beta_{\rm min}$
is the value of $\beta$ at which the average reflectance takes 
the minimum value. The dotted line
is a numerical fit: $\beta_{\rm min}=aC^b$ with $a\approx 2.04\pm 0.03$
and $b\approx 0.58\pm 0.01$.}
\end{figure}

\begin{figure}
\protect\centerline{\epsfxsize=5in \epsfbox{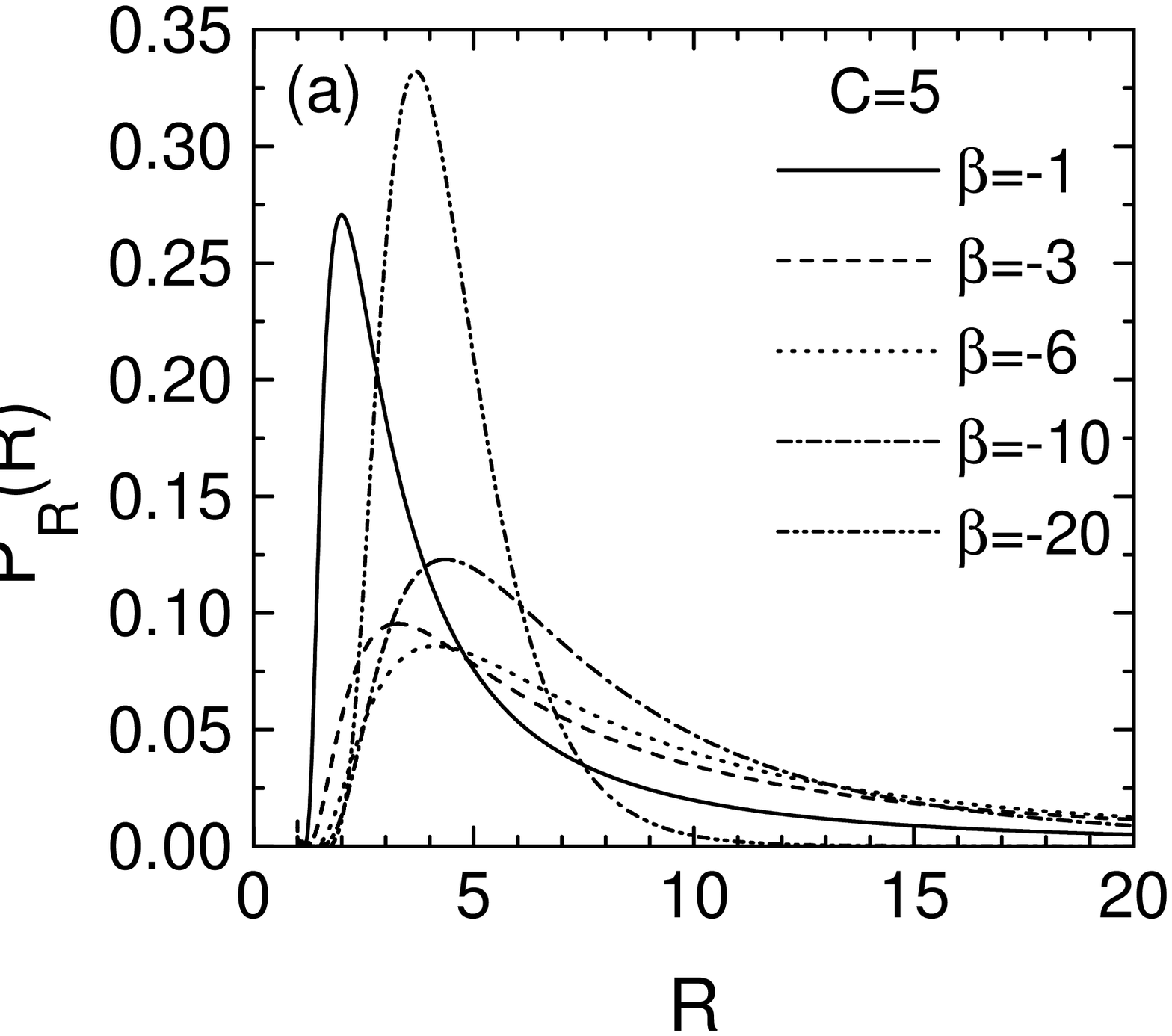}}
\protect\centerline{\epsfxsize=5in \epsfbox{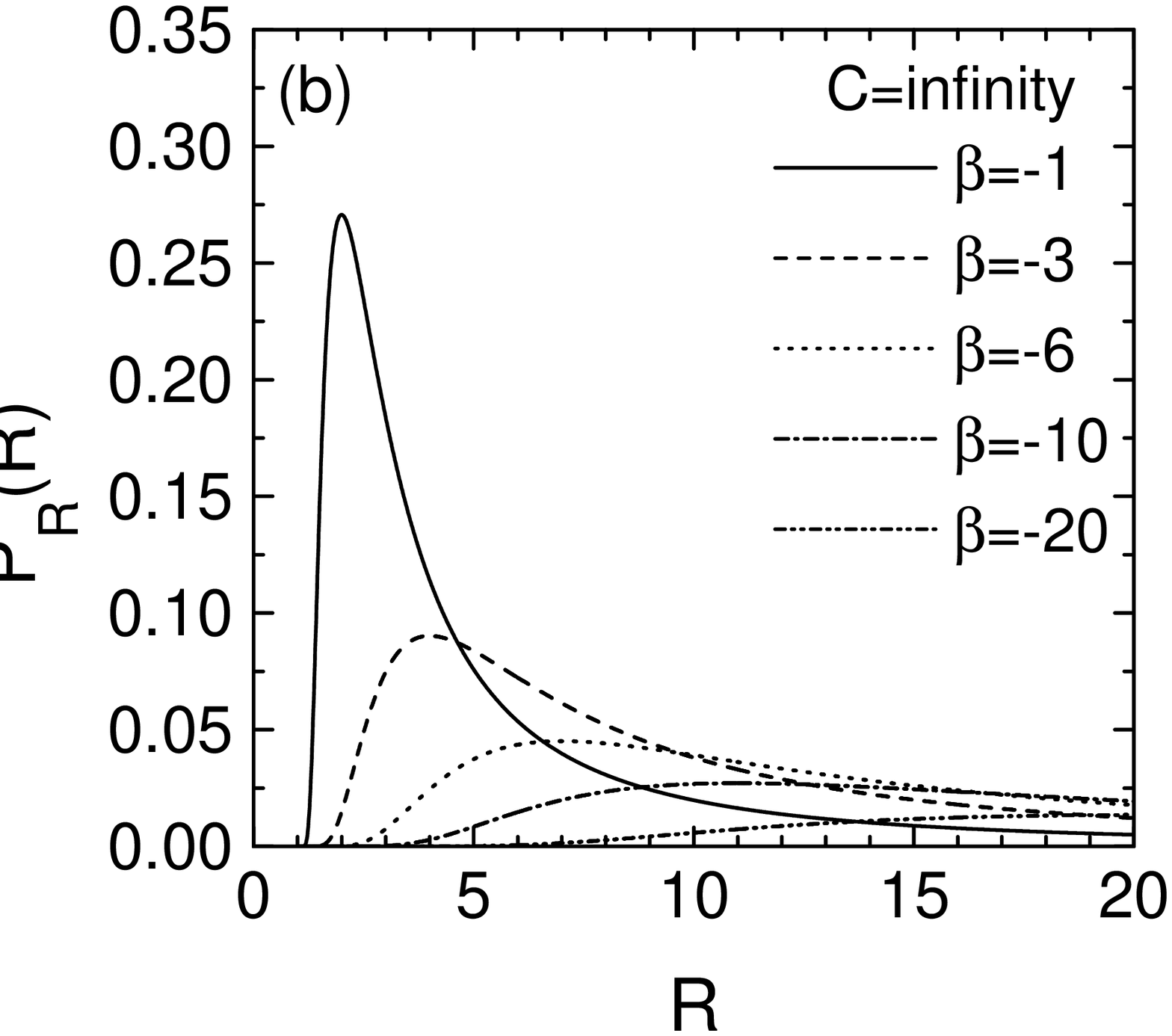}}
\caption{Probability density of the reflectance in the amplifying case
for $\alpha=0$, $\beta=-1,-3,-6,-10,-20$ and (a) $C=5$ (exact result) or  
(b) $C=\infty$ (RPA).}
\end{figure}

\begin{figure}
\protect\centerline{\epsfxsize=5in \epsfbox{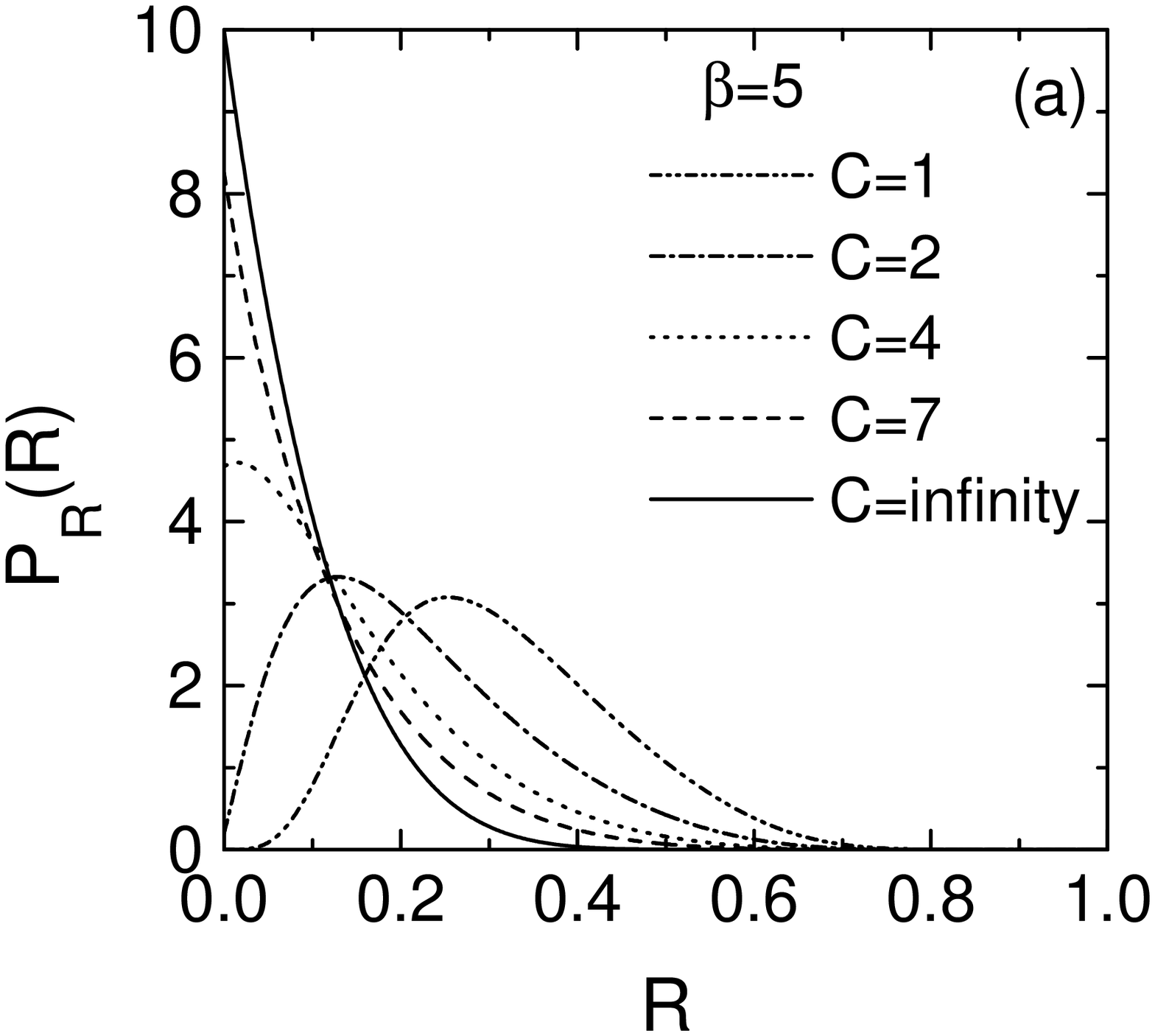}}
\protect\centerline{\epsfxsize=5in \epsfbox{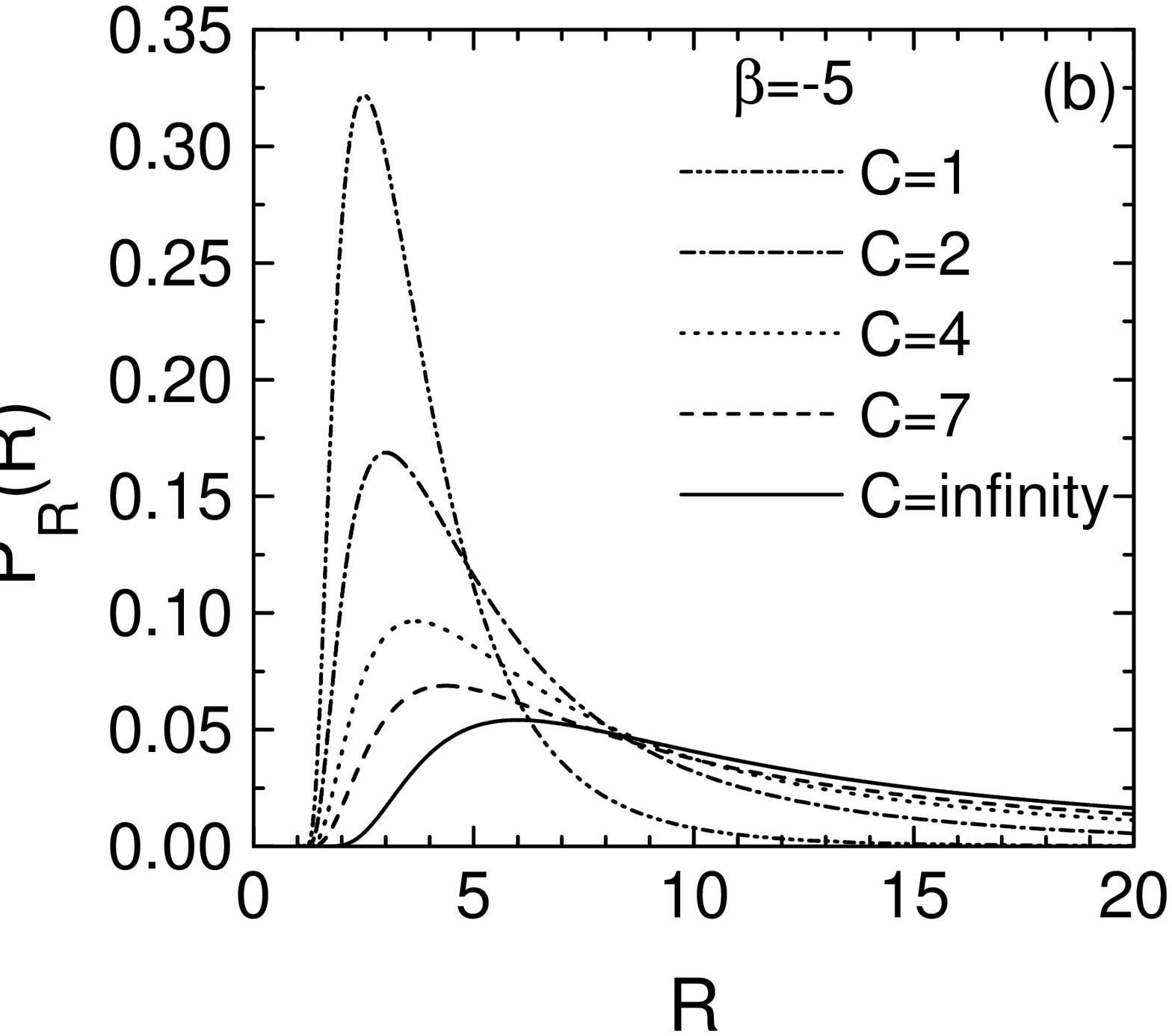}}
\caption{Probability density of the reflectance (a) in the 
absorbing case ($\beta=5$) and (b) in the amplifying case ($\beta=-5$)
for $\alpha=0$ and $C=1,2,4,7,\infty$.}
\end{figure}

\begin{figure}
\protect\centerline{\epsfxsize=5in \epsfbox{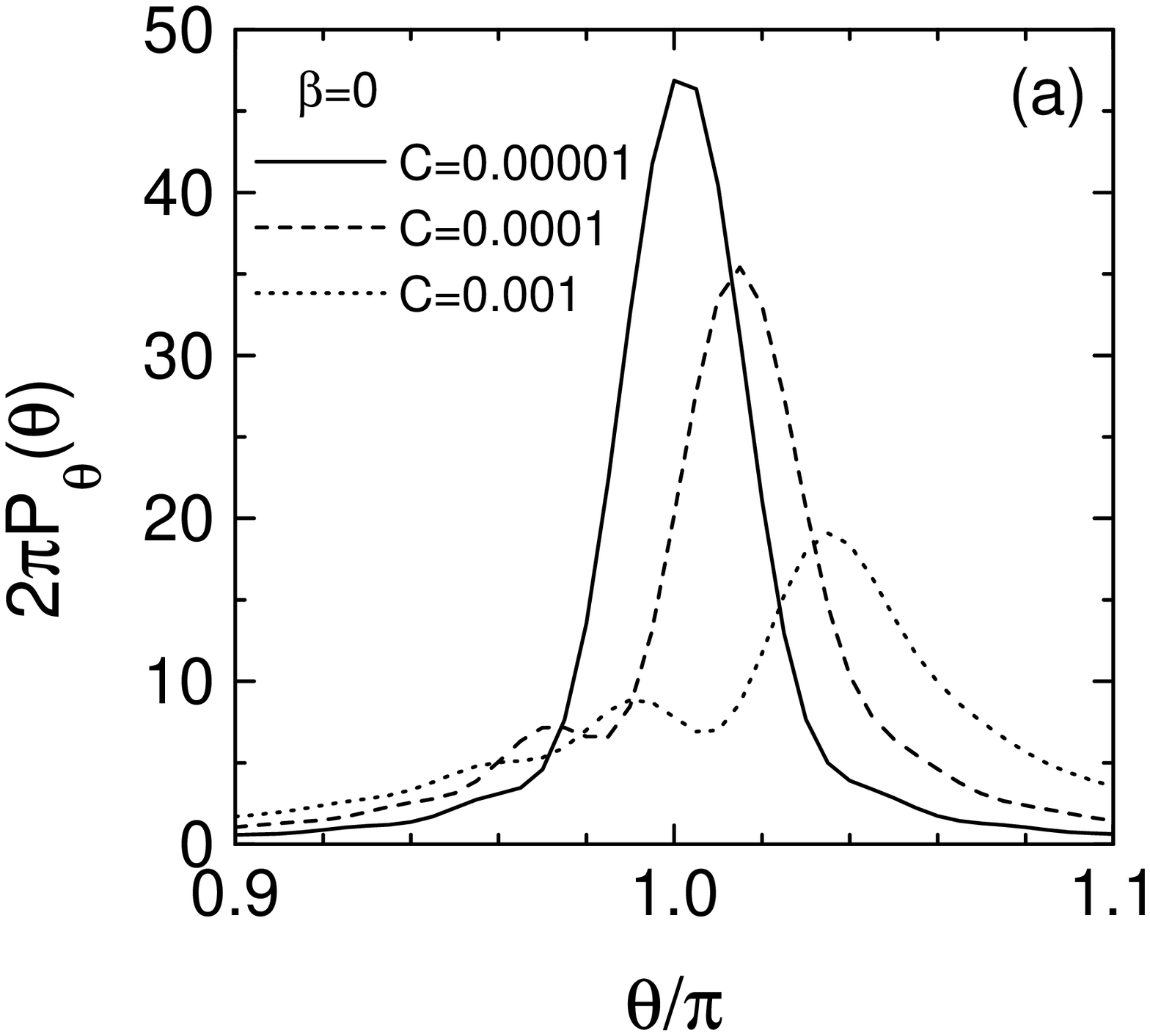}}
\protect\centerline{\epsfxsize=5in \epsfbox{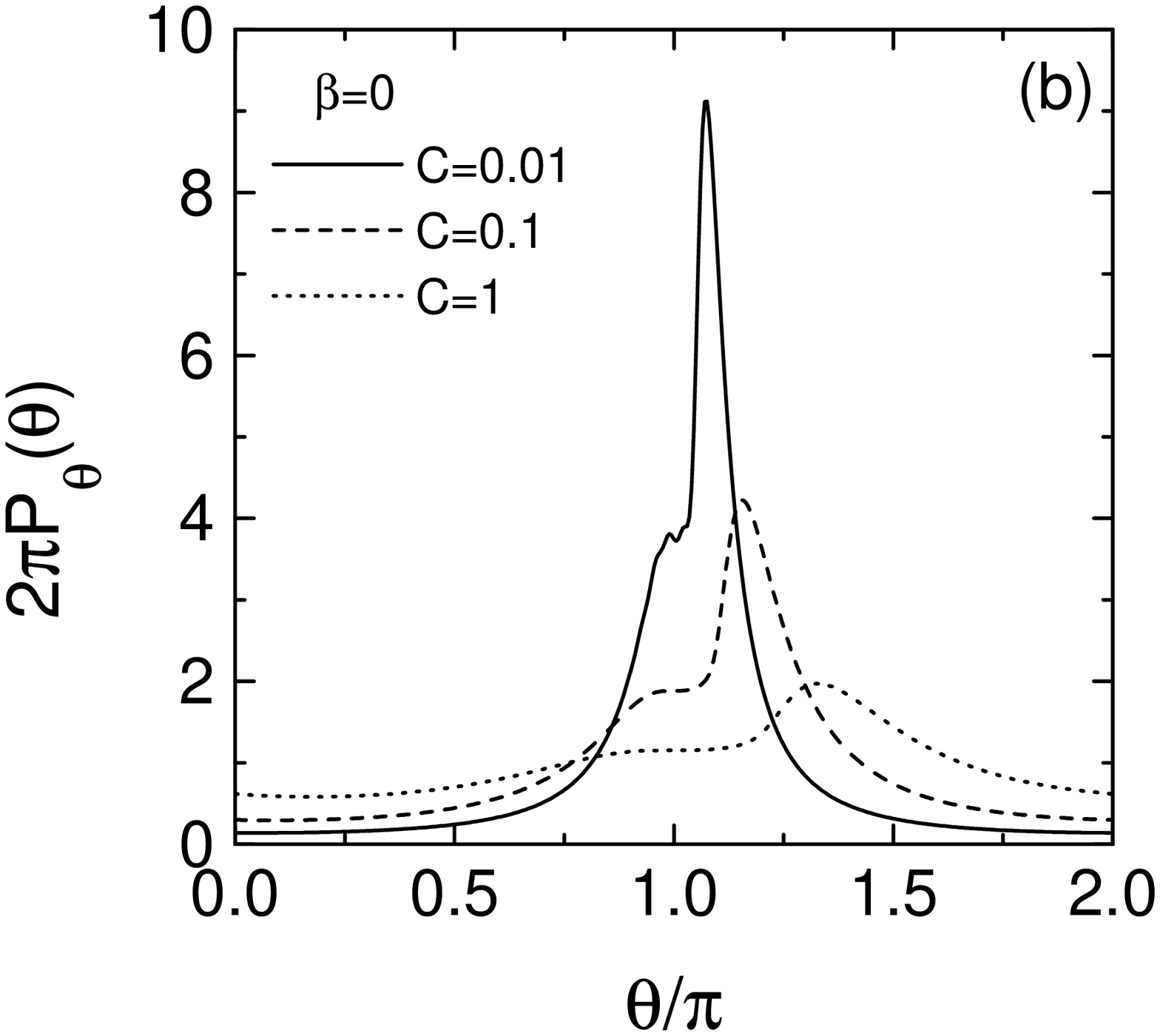}}
\protect\centerline{\epsfxsize=5in \epsfbox{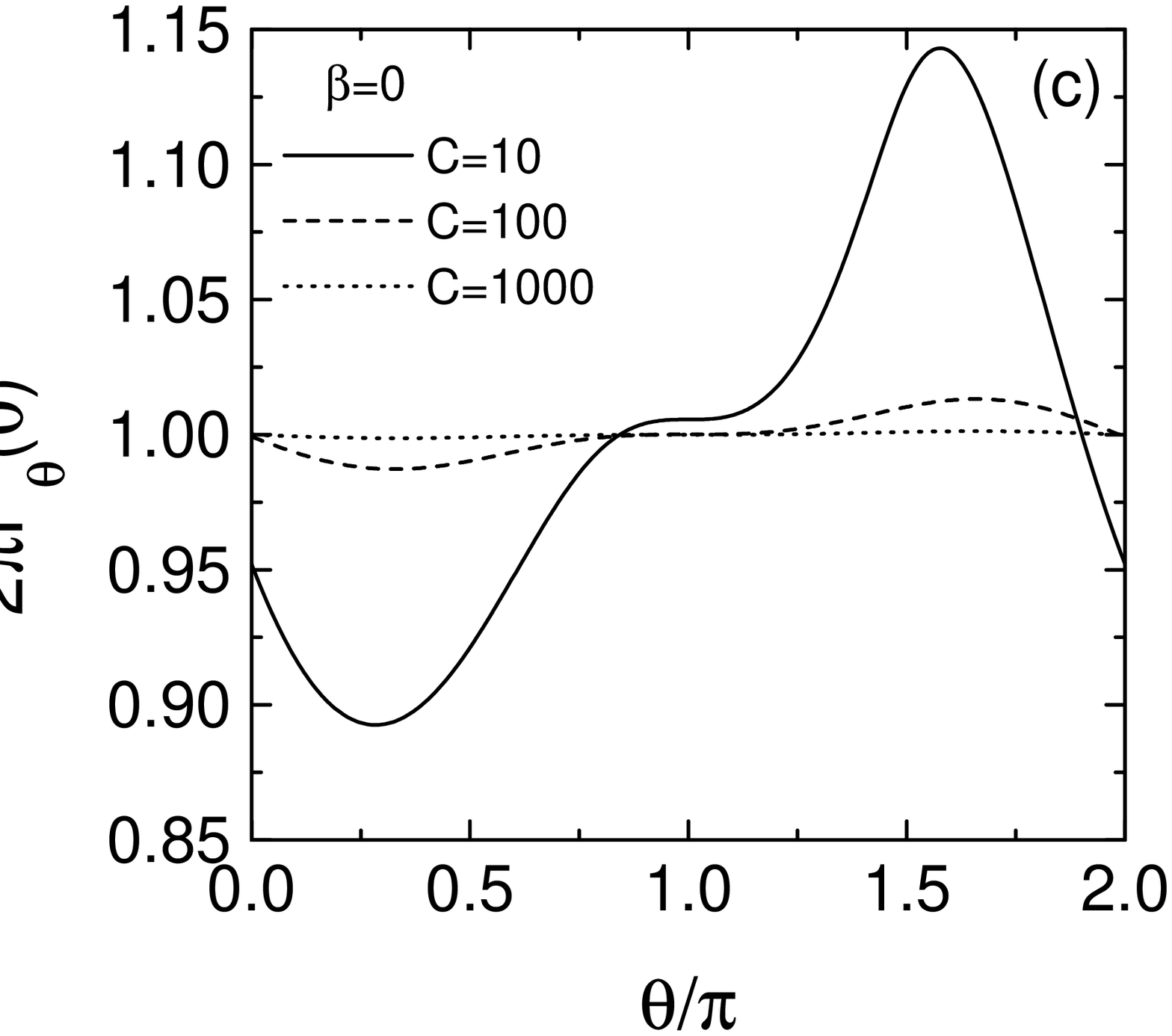}}
\caption{Probability density of the phase of the reflection coefficient
in the unitary case ($\beta=0$) for $\alpha=0$ and (a) $C=0.00001,
0.0001,0.001$ (b) $C=0.01, 0.1,1$ (c) $C=10,100,1000$.}
\end{figure}

\begin{figure}
\protect\centerline{\epsfxsize=5in \epsfbox{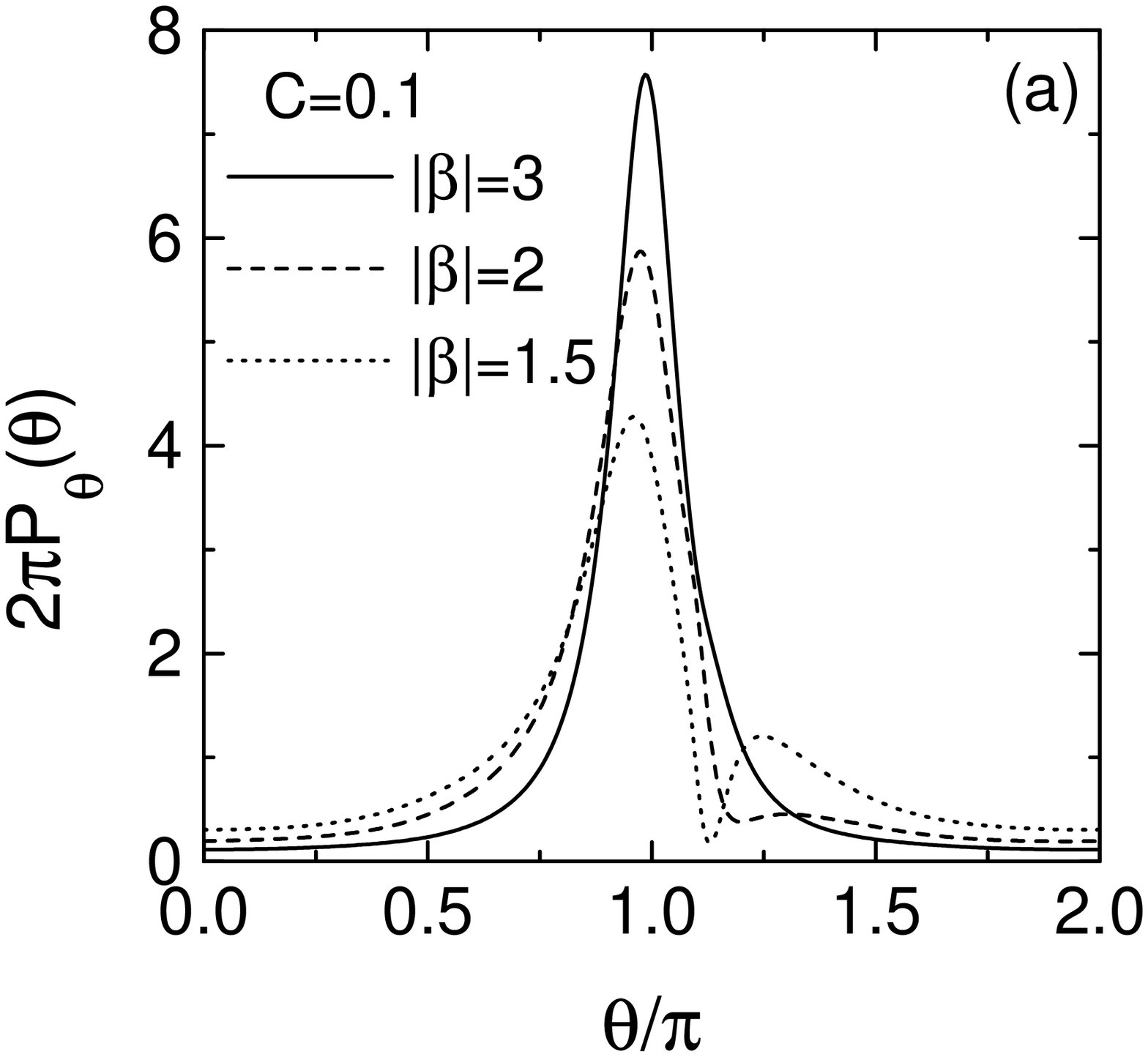}}
\protect\centerline{\epsfxsize=5in \epsfbox{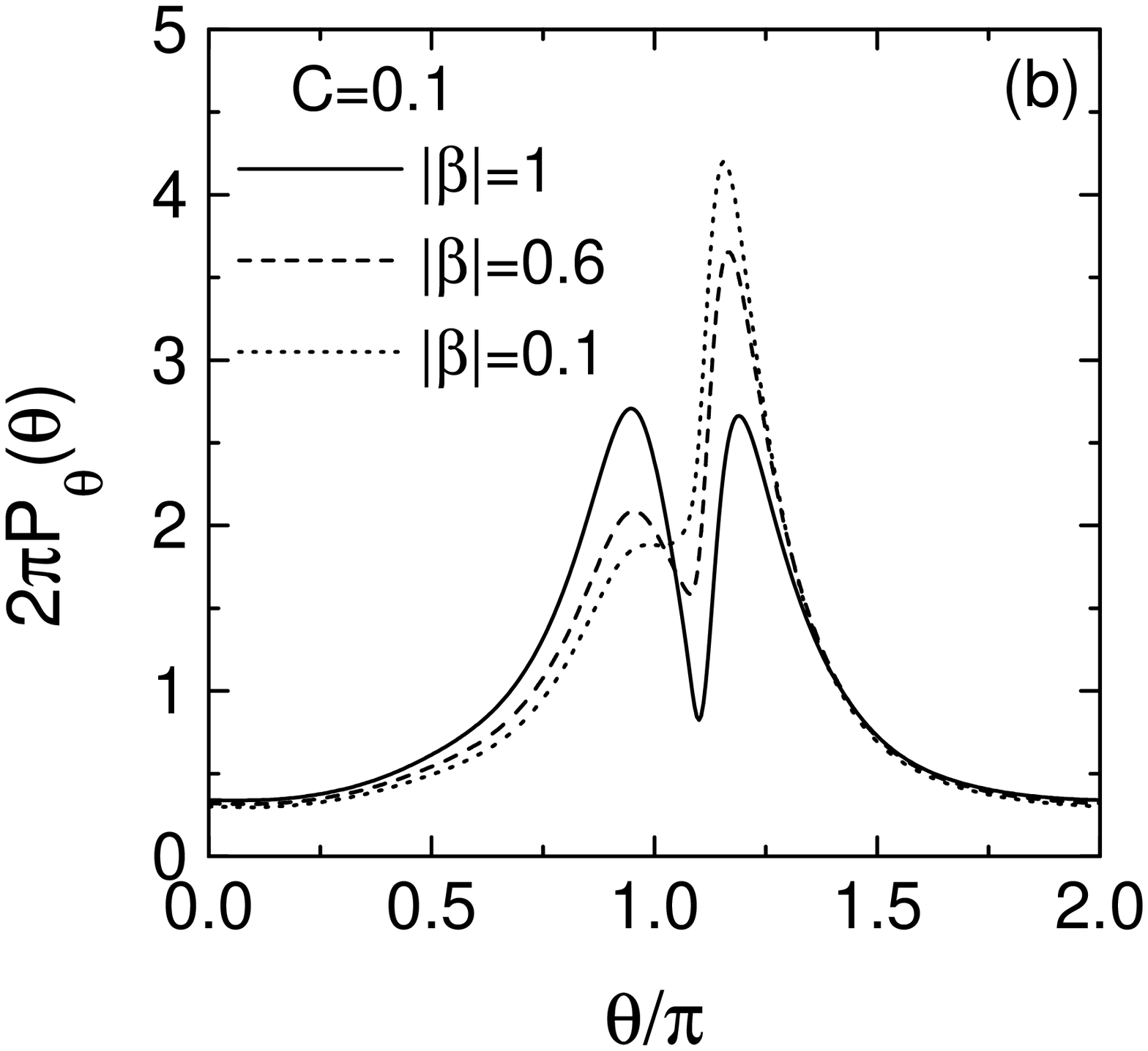}}
\caption{Probability density of the phase of 
the reflection coefficient, $P_\theta(\theta)$,
in the absorbing and amplifying cases for $C=0.1$, $\alpha=0$ and
(a) $\vert\beta\vert=3,2,1.5$
(b) $\vert\beta\vert=1,0.6,0.1$. Note that $P_\theta(\theta,\beta)
=P_\theta(\theta,-\beta)$.}
\end{figure}

\begin{figure}
\protect\centerline{\epsfxsize=5in \epsfbox{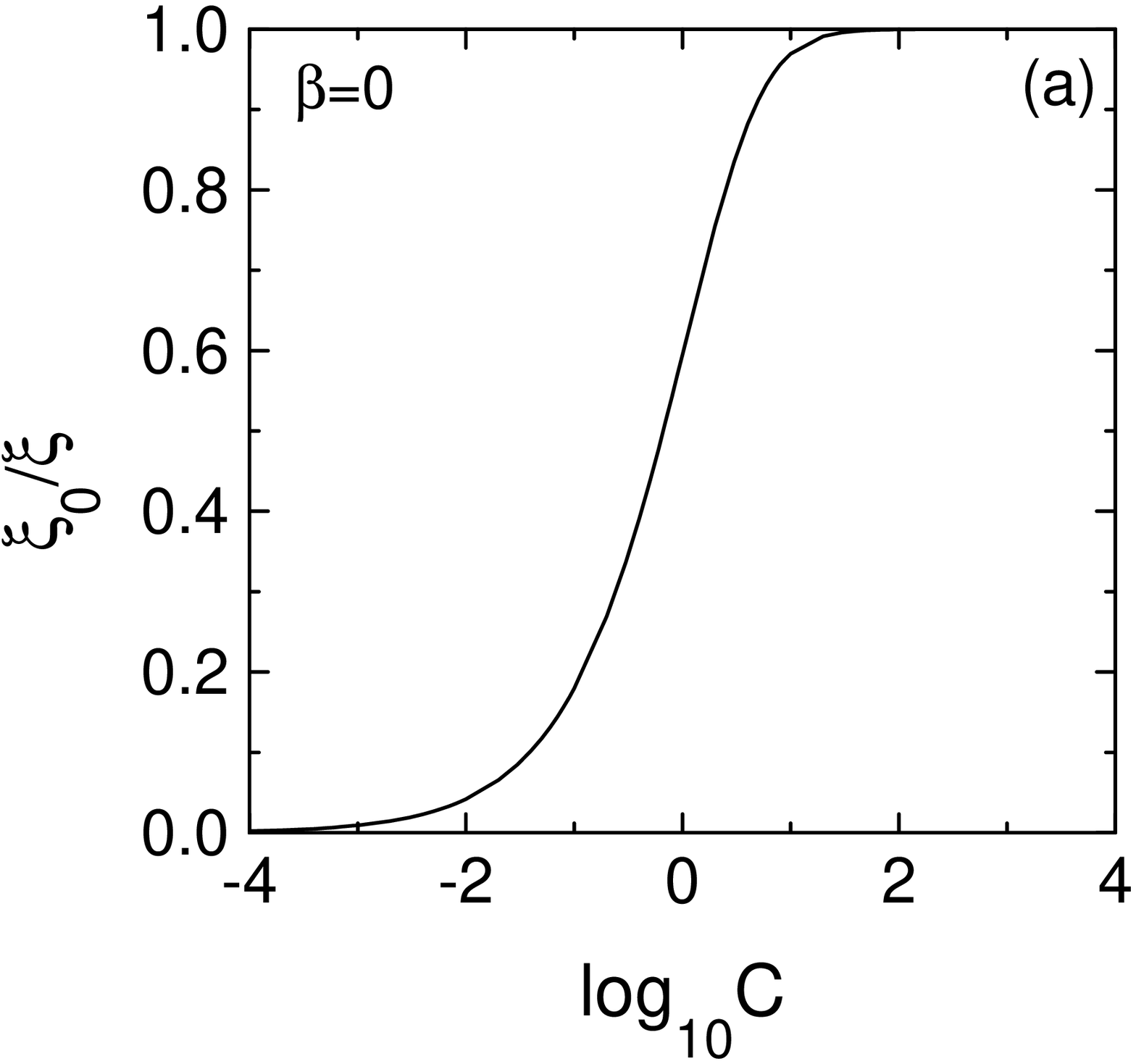}}
\protect\centerline{\epsfxsize=5in \epsfbox{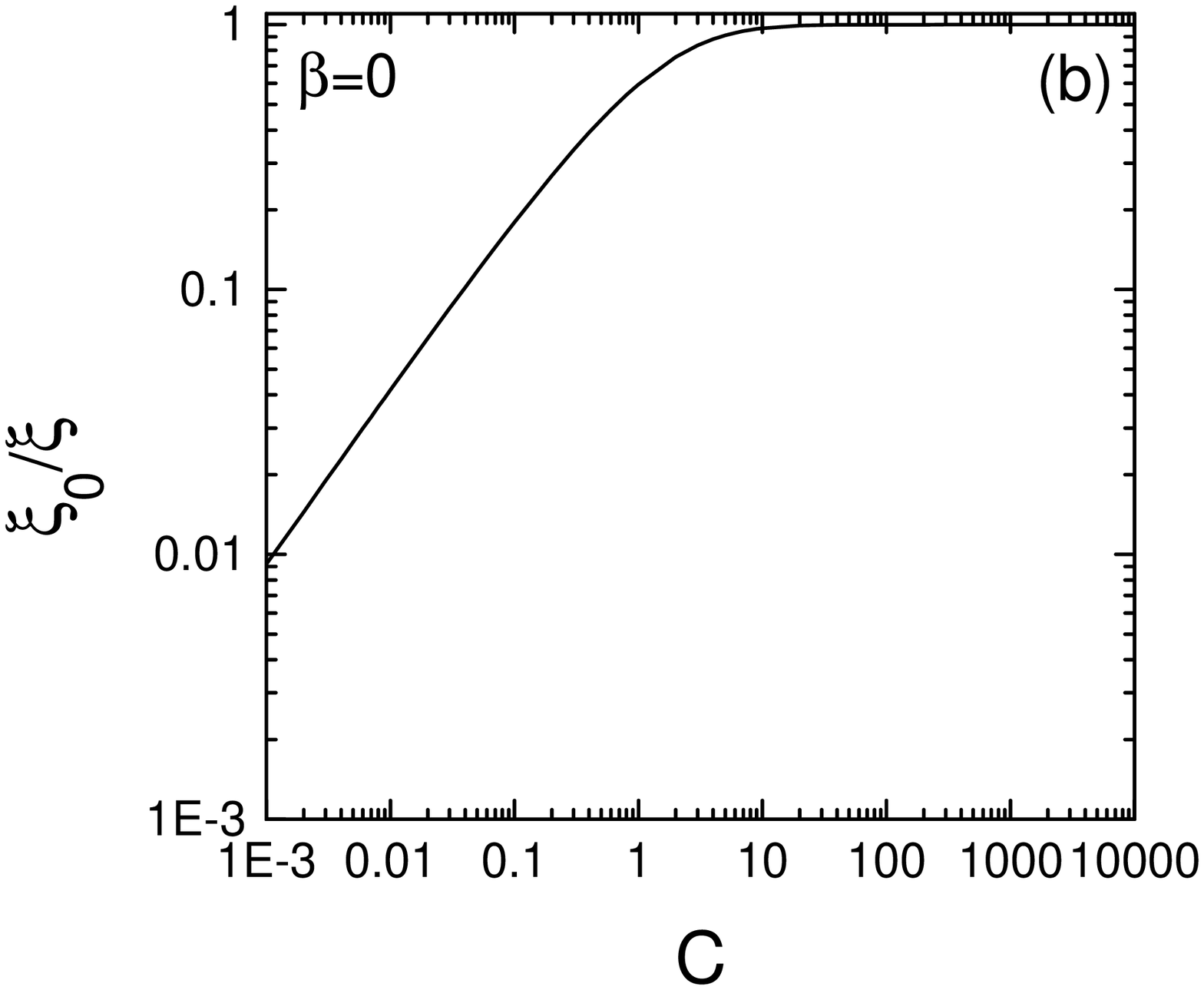}}
\caption{Inverse dimensionless localization length in the 
unitary case for $\alpha=0$ (a) versus $\log_{10}C$
 and (b) versus $C$ on a log-log plot. $\xi_0$ is the RPA localization
length in the unitary case.}
\end{figure}

\begin{figure}
\protect\centerline{\epsfxsize=5in \epsfbox{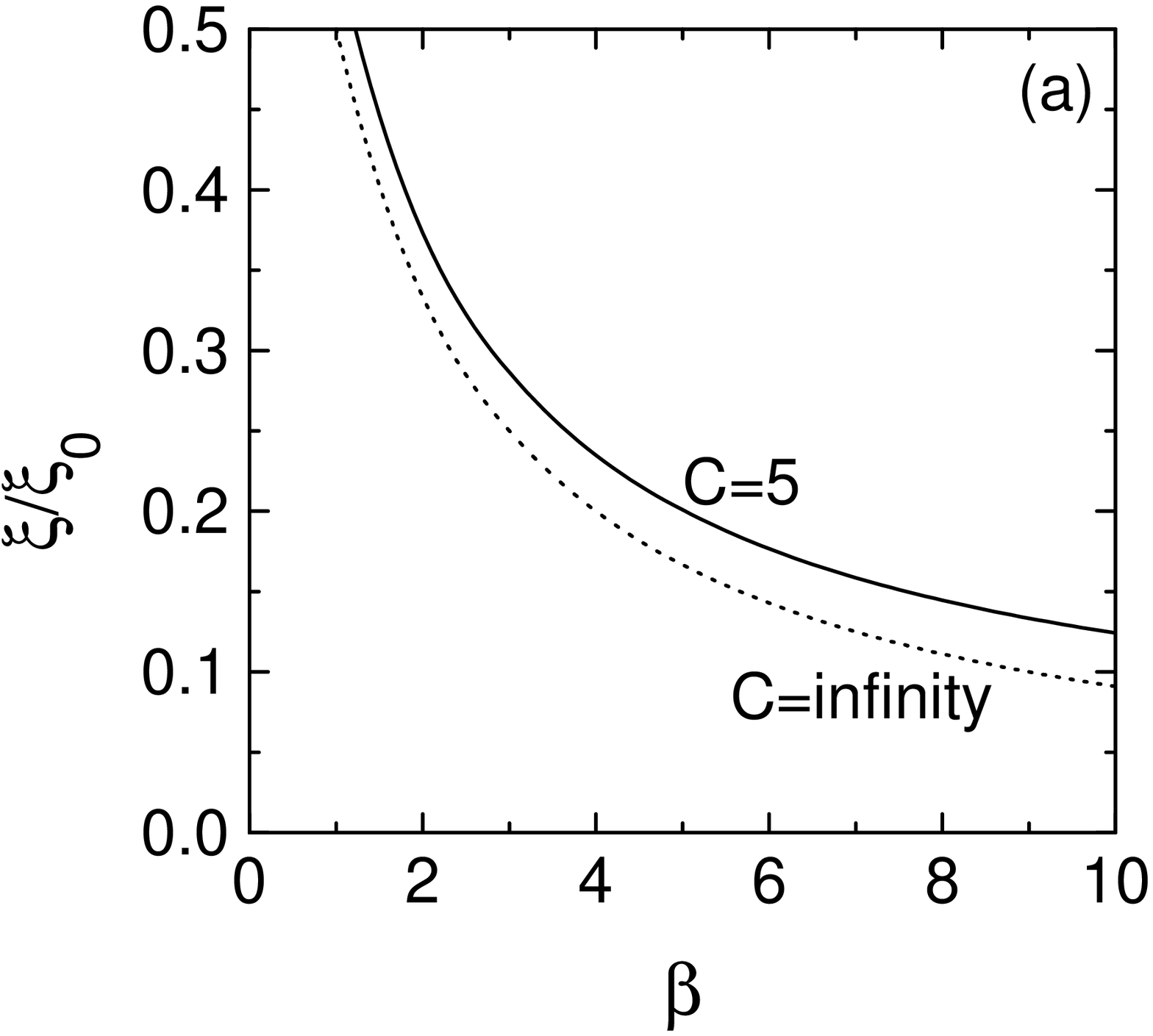}}
\protect\centerline{\epsfxsize=5in \epsfbox{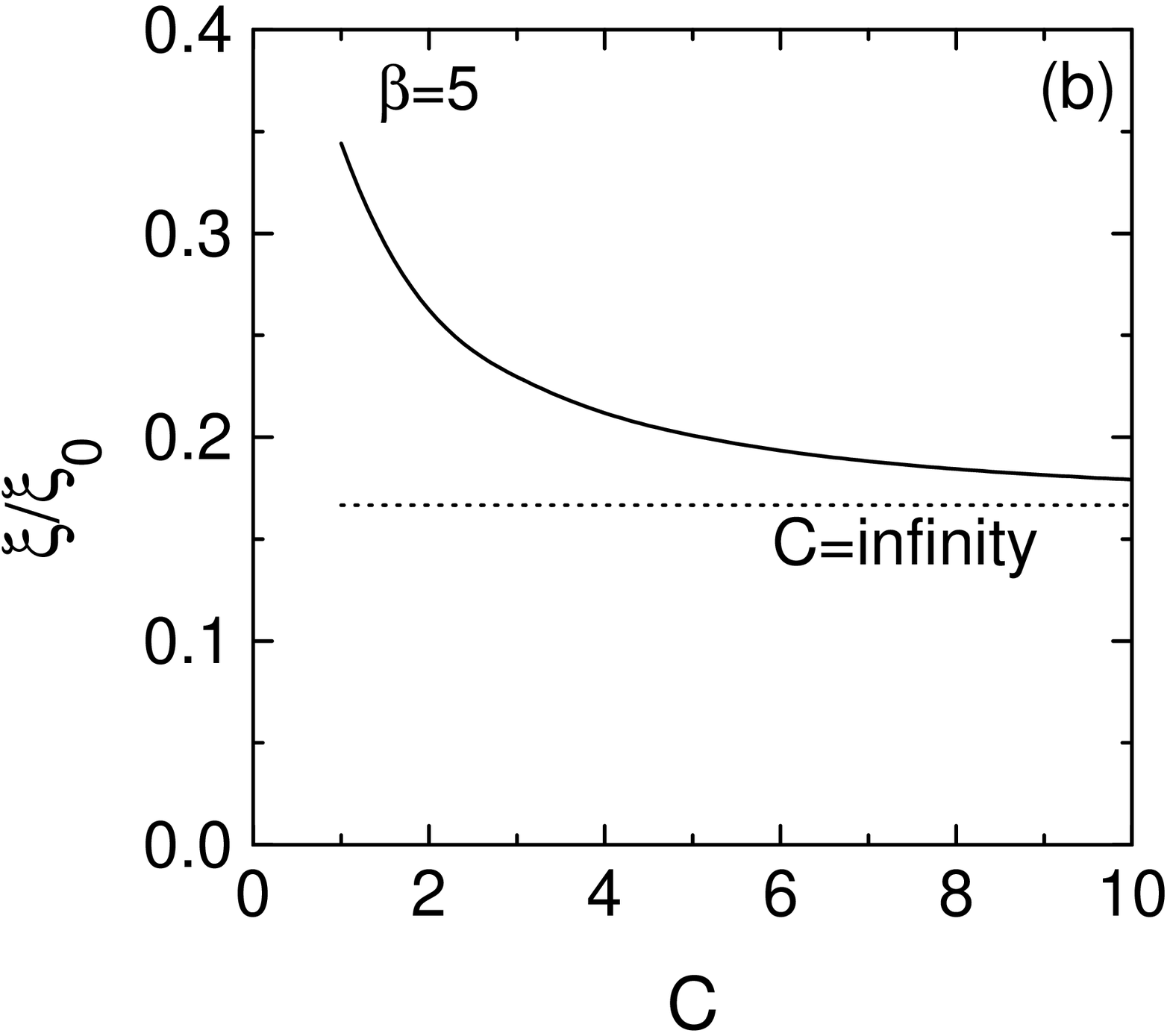}}
\caption{Dimensionless localization length in the absorbing case with $\alpha=0$
(a) versus $\beta$ for $C=5,\infty$ and (b) versus $C$ for $\beta=5$.
$\xi_0$ is the RPA localization
length in the unitary case.}
\end{figure}

\end{document}